# Modulating Super-Exchange Strength to Achieve Robust Ferromagnetic Couplings in Two-Dimensional Semiconductors

*Jiewen Xiao[1], Dominik Legut[2], Weidong Luo[3], Xiaopeng Liu[1], Ruifeng Zhang[1] and Qianfan Zhang[1*]*

1. School of Materials Science and Engineering, Beihang University, Beijing 100191, P. R. China.

2. IT4Innovations & Nanotechnology Centre, VSB-Technical University of Ostrava, 17.listopadu 2172/15, Ostrava CZ-70800, Czech Republic.

3. Institute of Natural Sciences, Shanghai Jiao Tong University, Shanghai 200240, P. R. China

*Corresponding author: qianfan@buaa.edu.cn.

**Abstract**

Low-dimensional semiconducting ferromagnets have attracted considerable attention due to their promising applications as nano-size spintronics. However, realizing robust ferromagnetic couplings that can survive at high temperature is restrained by two decisive factors: super-exchange couplings and anisotropy. Despite widely explored low-dimensional anisotropy, strengthening super-exchange couplings has rarely been investigated. Here, we found that ligands with lower electronegativity can strengthen ferromagnetic super-exchange couplings and further proposed the ligand modulation strategy to enhance the Curie temperature of low-dimensional ferromagnets. Based on the metallic $CrX_2$ (X = S, Se, Te) family, substituting ligand atoms by halides can form stable semiconducting phase as CrSeCl, CrSeBr and CrTeBr. It is interesting to discover that, the nearest ferromagnetic super-exchange couplings can be strengthened when substituting ligands from S to Se and Te. Such evolution originates from the enhanced electron hopping integral and reduced energy intervals between $d$ and $p$ orbits. While the second nearest anti-ferromagnetic couplings are also benefitted due to delocalized $p$-$p$ interactions. Finally, ligand modulation strategy is applied in other ferromagnetic monolayers, further verifying our theory and providing a fundamental understanding on controlling super-exchange couplings in low-dimension.

# I. INTRODUCTION

Two-dimensional (2D) materials have attracted tremendous interest in recent years and there are increasing 2D layered materials with unique physical and chemical properties that have been theoretically predicted and fabricated in experiments since the discovery of graphene.[1] Although great success has been achieved in 2D materials, realizing low dimensional ferromagnetism remains as a critical topic. Appealing 2D ferromagnets have promising applications in nano-size spintronic devices, which enables both the low energy consumption and high storage density. Besides, when forming heterojunctions with topological insulators, ferromagnetic monolayers can introduce magnetic proximity effect and break the time reversal symmetry, further opening up a gap in the surface states and realizing quantum anomalous Hall effect.[2][3] Nevertheless, long-range ferromagnetic order in low dimension has only been observed in few materials, and there are three famous monolayer phases, as $CrI_3$, $Cr_2Ge_2Te_6$, and $Fe_3GeTe_2$, have been received considerable attention.[4] Extensive investigated $CrI_3$ monolayer is semiconducting Ising ferromagnet, with out-of-plane easy axis. Although such anisotropy is remarkable, ferromagnetic couplings via super-exchange interaction among the Cr-I-Cr path is rather week, resulting in the Curie temperature as low as 45 K.[5][6] $Cr_2Ge_2Te_6$ is a Heisenberg magnet and the long-range ferromagnetic order can be established by external magnetic field, but is also limited by the weak ferromagnetic super-exchange couplings.[7] On the other aspect, $Fe_3GeTe_2$ monolayer exhibits metallic properties with iterant ferromagnetism, whose $T_C$ is around 68 K, but can be tuned to the room temperature through an ionic gate.[8][9] The iterant ferromagnetic exchange via carriers in $Fe_3GeTe_2$ is much stronger than super-exchange, but semiconducting ferromagnets would be more desirable due to their moderate band gaps and promising applications as transistors and spintronics. Therefore, it is necessary to further explore semiconducting magnets with robust ferromagnetic couplings that can survive at high temperature.

However, the realization of strong low-dimensional ferromagnets depends on two decisive factors: anisotropy and super-exchange couplings. For the first anisotropy requirement, Mermin-Wagner theorem regulates that long-range magnetic order cannot exist in the isotropic two-dimensional system.[10] Nevertheless, only a small anisotropy is enough to open up a sizable gap in the magnon spectra, thus stabilizing magnetic orders against finite temperature.[11] Experiments on $Cr_2Ge_2Te_6$ indicate that external magnetic field can also introduce anisotropy, and such strategy is promising to be applied in other systems only if they are not strong XY magnets. On the other aspect, super-exchange theory has long been established by Goodenough, Kanamori and Anderson (GKA), which mainly focuses on ionic compounds with oxygen as ligand.[12-15] While in low dimension, large ligands as S/Se/Te/Br/I can rather stabilize the monolayer structure. However, ligands other than oxygen will lead to the enhanced covalency and further deviate from the ionic picture, and the modification of super-exchange couplings via controlling the degree of *d-p* hopping process can thus be realized. Furthermore, substituting ligands has already been achieved in the MoSSe system,[16][17] which suggests the feasibility of this ligands modulation strategy. However, the exploration on the role of ligands and their effect on magnetic properties have rarely been studied, but is rather essential for the modulation on the strength of exchange couplings, especially in low-dimension.

In the present work, we have applied ligand modulation strategy on the metallic 1-T $CrX_2$ (X = S, Se, Te) family, since Cr based compounds are known to exbibit ferromagnetism in low-dimension. Through multiple screening rules, CrSeBr, CrSeCl and CrTeBr were selected as the semiconducting Janus monolayer with the robust ferromagnetic order, while CrTeCl adopts antiferromagnetic couplings. To understand their magnetic properties, we concentrate on exchange integrals, and the nearest exchange couplings can be determined by three kinds of super-exchange process. It is discovered that, varying ligands from S to Se and Te can enhance the electron hopping integral and reduce energy intervals between $d$ and $p$ orbits at the same time, thus strengthening ferromagnetic couplings. Further calculation reveals that the second nearest exchange integral can also be significantly affected by ligands, where $p$-$p$ hopping process in telluride compounds is much more effective due to the delocalized feature of $p$ electrons. Finally, we examined our theory on a series of reported low-dimensional magnets, further demonstrating our discoveries. Our work not only reveals a new family of low-dimensional magnets with mixed ligands, but also provides a fundamental understanding on the modulation of super-exchange couplings.

## II. COMPUTATIONAL METHODS

First-principles calculation has been performed in the framework of density functional theory using the Vienna Ab initio Simulation Package (VASP).[18][19] Generalized gradient approximation (GGA) exchange-correlation was described by the Perder-Burke-Ernzerhof (PBE) formulation.[20] The projector augmented wave (PAW) pseudopotentials[21][22] was adopted to describe the interaction between electrons and nuclei. An energy cutoff 500eV was employed for the plane wave basis. The criteria of the total energy convergence and the atomic force tolerance was set to $10^{-5}$ eV and 0.01 eV/ Å respectively. For describing the Fermi-Dirac distribution function, a Gaussian smearing of 0.02eV was used. The 13×13×1 Gamma-centered Monkhorst-Pack grids[23] was employed to sample the Brillouin zone for the relaxation of all structures. While for electronic structure calculation, denser grids were set to 24×24×3. Considering the localized nature of 3d electrons for transition metals, the DFT+U method was adopted.[24] The effective U-J value was tested ranging from 1 to 5 eV, and for magnetic and electronic calculations, the U-J value was set to 2.8eV in accordance with the previous work.[25][26] To describe the electronic structure more precisely, the HSE06 (Heyd-Scuseria-Ernzerhof) hybrid functional[27][28] was further applied. To analyze bonding and anti-bonding states in detail, Crystal Occupation Hamilton Population (COHP) was applied through using LOBSTER.[29-32]

In order to validate that the Janus monolayer structure for CrSeX, CrTeX (X= Br, Cl) are with the lowest energy, the structure prediction was performed using the *ab-initio* random structure searching method (CALYPSO) based on particle swarm optimization (PSO) algorithm.[33][34] The Janus structure was reproduced among more than 10,000 generated structures and the dynamical stability was verified based on the phonon spectrum simulated by PHONONPY. In order to further test the stability at room temperature, the *ab-initio* molecular dynamics (AIMD)[35][36] simulation was applied using the Nose heat bath scheme. The canonical ensemble at 300K was adopted to simulate the thermal stability for the 4×4 supercell of CrXY (X = Se, Te; Y= Cl, Br).

## III. RESULTS AND DISCUSSIONS

### A. Structure and stability

Based on the metallic 1-T phase of $CrX_2$ (X = S, Se, Te), we have replaced one layer of S/Se/Te by Cl/Br/I and further screen those candidates through comparing the total energies of a series of common 2D structures with the same atomic ratio (Figure S1). CrSeCl, CrSeBr, CrTeCl and CrTeBr are selected out, which retain the 1-T phase as the most stable atomic configuration (Table S1). Structure prediction algorithm is additionally performed, further verifying that their ground states as the 1-T Janus monolayer. Figure 1a shows the atomic configuration of CrSeBr while others are presented in Figure S2. Four candidates possess hexagonal symmetry belonging to the space group 156. And the inversion symmetry is broken due to the replacement of one layer of original ligand atoms by halides. Among this series, the bond angles for Cr-Se/Te/Cl/Br-Cr are all around 90°, suggesting a favored ferromagnetic super-exchange couplings based on the GKA rule.[37][38] Lattice dynamics (phonon dispersion relations) were further calculated shown in Figures 1b and S2. The lack of imaginary modes demonstrates their excellent dynamical stability. Corresponding phonon density of states is presented in the right panel of Figure 1b, indicating that three higher optical modes are mainly contributed by the correlative vibration of Cr and Se atoms while other three lower lying optical bands correspond to Se and Br atoms. *Ab-initio* molecular dynamics simulation was additionally performed, which shows that the honeycomb network remains intact after simulating 20 *ps* at 300 K, validating their thermodynamic stabilities at room temperature (Figure S3).

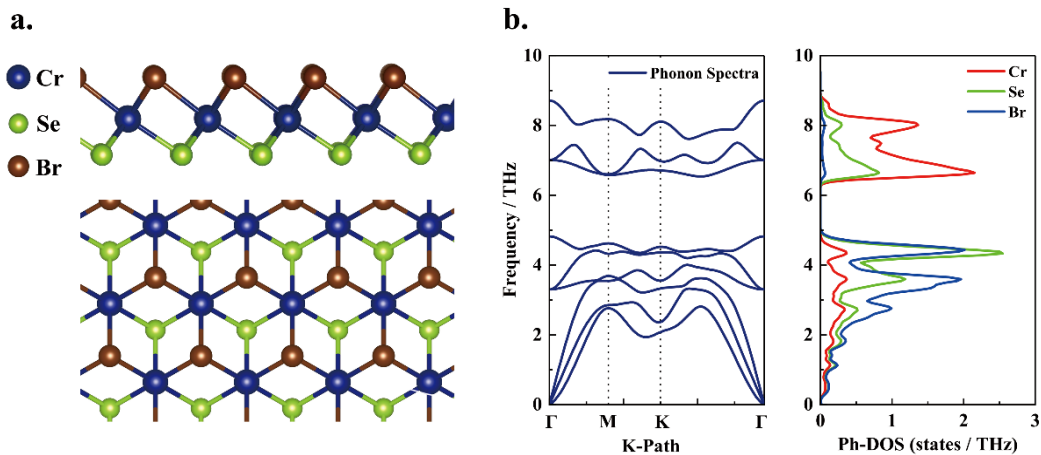

FIG. 1. a. Top view and side views of atomic configuration for CrSeBr; b. Phonon dispersion relation and phonon density of states (Ph-DOS) for CrSeBr.

**B. Magnetic and electronic properties**

Magnetic properties are further calculated, and magnetic ground states are firstly derived by comparing total energies of different magnetic configurations (Figure S4). It is found that CrSeCl, CrSeBr and CrTeBr all favor ferromagnetic couplings, and the energy of FM order is ~ 40 meV per magnetic cation lower than that of the stripy AFM order. CrTeCl, in contrast, prefers the stripy AFM order by ~ 3 meV per magnetic cation. For three ferromagnetic candidates, the overall magnetic moment of a single unit cell is 3 $\mu_B$. While local magnetic moments of Cr atoms are calculated to be ~ 3.3 $\mu_B$, based on both PBE+U and HSE06 scheme (Table 1), suggesting the rationality of our adopted U value. It is worth to note that there is a sizable induced magnetic moment on ligands, especially for Se and Te with the value around -0.3 $\mu_B$, being similar the spin-polarized iodine in the $CrI_3$ system. Magnetic anisotropy energy (MAE) induced by spin-orbit coupling (SOC) is further calculated, where MAE is defined as $E_{out-of-plane} - E_{in-plane}$ per unit cell. As shown in Table 1, the existence of easy-plane indicates that CrSeCl, CrSeBr are weak XY magnets while CrTeBr is relatively strong. To restore the long-range magnetic order in finite temperature, we have briefly explored two strategies to introduce anisotropy. The first strategy is forming heterojunction with anisotropic $CrI_3$ monolayer, and such ferromagnetic substrate resembles the applied the external magnetic field. And, for weak XY magnets as CrSeCl and CrSeBr, the easy-axis can be restored in the whole heterojunction system, and the strengthen of it is similar with the remarkable value for $CrI_3$. The second strategy is about uniaxial strain engineering, and we have applied a series of in-plane uniaxial strain and further calculated the value of MAE (Figure S5 and Table S2). Results indicate that uniaxial strain can break the in-plane isotropy, and such geometric anisotropy will correlate with spin directions via SOC, whose magnitude is small but can also open the gap in magnon spectra and further repress quantum fluctuations. Finally, for the CrTeCl with the stripy AFM order, it exhibits the strong in-plane anisotropy (~ 1656 μeV) with the easy axis along the [100] direction.

Next, since the magnetic anisotropy is small, we quantitatively describe the magnetic exchange interactions based on the Heisenberg spin Hamiltonian:

$$H = -J_1 \sum_{<ij>} \vec{S_i} \cdot \vec{S_j} - J_2 \sum_{<<ij>>} \vec{S_i} \cdot \vec{S_j}$$

Where <ij> and <<ij>> represent the first and second nearest couplings, which is generally sufficient to describe the exchange couplings in magnetic monolayers.[39][40] The calculated exchange parameters as $J_1$ and $J_2$ are listed in Table 1. Generally, this series exhibit ferromagnetic $J_1$ ~ 30 meV, and it is almost ten times of the value for $CrI_3$.[41] While the second nearest exchange interaction $J_2$ favors AFM order, which is significantly larger for telluride compounds. To further identify whether there exists a general evolution trend of $J_1$ and $J_2$ in these compounds, we further explored other CrXY (X = S/Se/Te, Y = Cl/Br/I) candidates, although they are metastable phases. As presented in Table S3-S5, it is interesting to discover that $J_1$ gradually increases when involving from S to Se and Te, while $J_2$ decreases and will reach an especially large value for telluride compounds.

TABLE I. Magnetic moment of Cr based on PBE+U and HSE06 scheme, exchange integrals ($J_1$ and $J_2$), magnetic anisotropy energy (MAE) for monolayer phases and heterojunctions.

| Properties | CrSeCl | CrSeBr | CrTeBr | CrTeCl |
|---|---|---|---|---|
| Magnetic moment (PBE+U)/ $\mu_B$ | 3.32 | 3.34 | 3.39 | 3.49 |
| Magnetic moment (HSE06)/ $\mu_B$ | 3.22 | 3.24 | 3.48 | 3.38 |
| $J_1$/meV | 28.23 | 29.48 | 34.94 | 33.49 |
| $J_2$/meV | -5.51 | -4.53 | -14.83 | -17.83 |
| MAE (Monolayer)/$\mu eV$ | 114 | 56 | 1298 | 1162 |
| MAE (Heterojunction)/ $\mu eV$ | 1375 | 878 | -628 | |

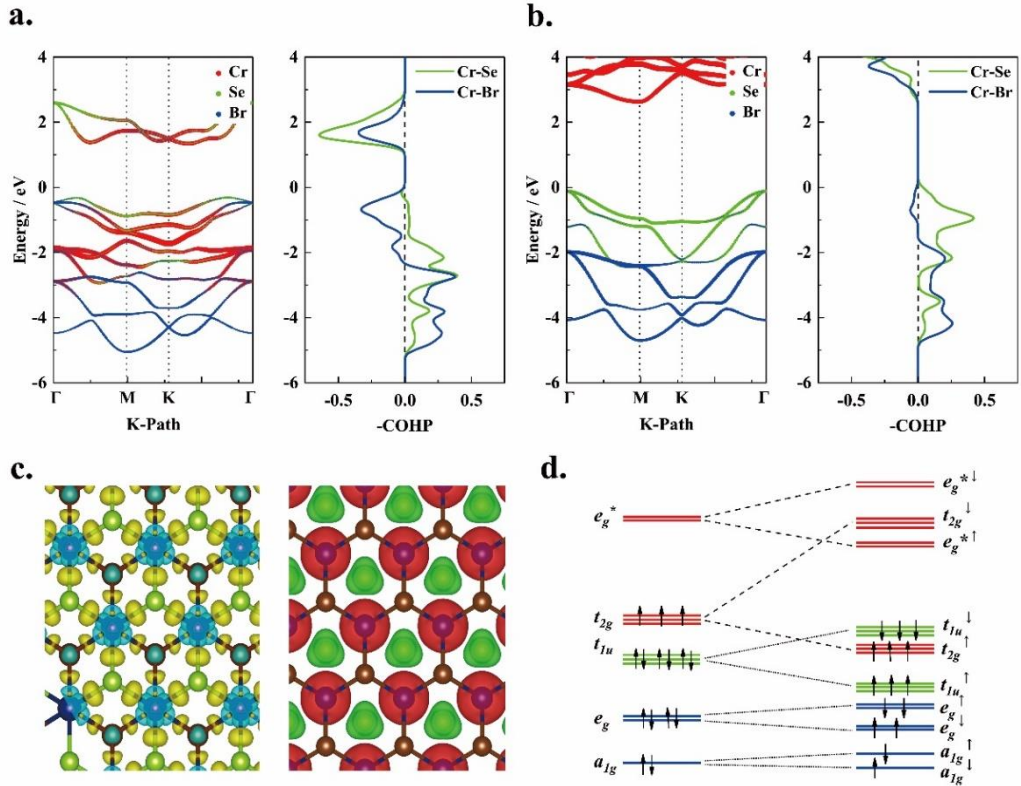

FIG. 2. Projected band structure and corresponding COHP for a. spin-up and b. spin-down channel in CrSeBr, where red, green and blue dots represent projected electronic states in Cr, Se and Br atoms. c. Differential charge density for CrSeBr (left panel),

where yellow and green region denotes charge accumulation and depletion with iso-surface being set to 0.01 e/Å. Spin density for CrSeBr (right panel), where red and green region represent two kinds of spin components, with iso-surface as 0.01 e/Å$^3$. d. Schematic representation of the evolution of electronic states based on ligand field theory, where red, green and blue energy levels denote those electronic states that are mainly occupied by Cr, Se and Br atoms respectively.

To further explore magnetic couplings, we firstly presented the electronic structure of CrSeBr in Figure 2a and 2b based on DFT+U scheme, as a typical representative. More precise HSE06 hybrid functional scheme is also performed, and band structures are presented in Figure S6. For both simulation methods, the band structure exhibits the similar semiconducting nature with indirect moderate band gap as 1.46 eV (DFT+U) and 2.43 eV (HSE06). Band structures for other monolayers are further shown in Figure S6, sharing the similar features. Atomic differential charge density along with the spin density is then plotted in Figure 2c. And it shows that, charge depletes around Cr atoms and accumulates in the region of Cr-Br/Se bonds, while Cr atoms are still responsible for the large spin polarization. In combination with the projected band structures (Figure 2a and 2b) and ligand field theory, the splitting and rearrangement of electronic states can be clarified and illustrated as Figure 2b. Generally, in CrXY (X = S/Se/Te, Y = Cl/Br/I) compounds, the valance state for Cr is +3, indicating its $d^2sp^3$ hybridization feature with the formation of 6 bonding and anti-bonding pairs, which is a typical case for the octahedral ligand field. Therefore, the six hybridized bonding states, corresponding to $a_{1g}$, twofold $e_g$ and threefold $t_{1u}$ orbits with different symmetries, are filled by 12 electrons: 9 from the $p$ orbits of Se/Br atom and 3 from $s/d_{x^2-y^2}/d_{xy}$ orbits of Cr atom. The remained non-bonding $t_{2g}$ orbits, composed of $d_{xz}/d_{yz}/d_{z^2}$, are half filled by three left electrons with spin-up component according to Hund's rule, representing a stable electronic configuration. Such bonding states can be demonstrated by the Crystal Occupation Hamilton Population (COHP) analysis. As shown in the right panel of Figure 2a and 2b, the six lower lying bands, composed of $a_{1g}$, twofold $e_g$ and threefold $t_{1u}$, correspond to the COHP bonding peaks for both Cr-Se and Cr-Br bonds. While $e_g^*$ orbits above Fermi level exhibit strong anti-bonding feature, being consistent with the former analysis. It is worth to note that, for threefold nonbonding $t_{2g}$ states right below Fermi level, moderate anti-bonding peaks will appear. We can identify such anti-bonding nature in $t_{2g}$ states as the consequence of super-exchange process in the following text. Next, after taking the exchange field into account, the spin-up and spin-down states will split. As shown in Figure 2b, $t_{2g}^{\downarrow}$ states are shifted far above Fermi level, while occupied $t_{2g}^{\uparrow}$ states sit right below empty $e_g^{*\uparrow}$ states, further being followed by $t_{1u}$ states mainly composed of $p$ orbits.

### C. Super-exchange mechanism and evolution of exchange integrals

Based on electronic structures, magnetic exchange mechanism can be elaborated. For semiconducting ferromagnetic CrSeBr series, there are two major kinds of exchange couplings: 1) direct exchange between magnetic cations; 2) super-exchange (SE) via $p$ orbits of ligands. Firstly, direct exchange is originated from the overlapping among two cations' non-orthogonal states, thus being AFM and sensitive to distances.[42] Since magnetic cations are separated by ligands, their interaction is rather week for 3d elements. Only heavy transition metals need to further consider their metallic interactions.[43] Next, for the second super-exchange process, it is responsible for the strong ferromagnetic couplings. In these Janus monolayers, geometry allows $Cr_1$-Se-$Cr_2$ and $Cr_1$-Br-$Cr_2$ bind with each other at a right angle, and there are three kinds of super-exchange scenarios based on different involved orbital symmetries, as shown in Figure 3a. The first mechanism (SE1) can be expressed as $t_{2g}$-$p_x/p_y$-$t_{2g}$, which means that, the $t_{2g}$ states in $Cr_1$ and $Cr_2$ will form π bonds with $p_x$ and $p_y$ states in the same ligand. Such π bond is described as the partial covalent bond by Goodenough,[13] which will allow spin-down electrons in the ligand to hop into $t_{2g}$ orbits. While spin-up electrons left in the $p_x/p_y$ state can ferromagnetically exchange with each other based on Hund's rule for such onsite orthogonal orbits. Second super-exchange process (SE2) can be presented as $e_g^*$-$p_x/p_y$-$e_g^*$, where electron hopping happens in $e_g^*$-$p_{x/y}$ via partial covalent σ bond, and ferromagnetic order is further maintained by the onsite orthogonal $p_x$ and $p_y$ exchange. Two former super-exchange process can be quantitatively expressed as:[38]

$$J_{t_{2g}/e_g-t_{2g}/e_g}^{SE1/2} \sim -\frac{t_{pdm}^2 t_{pdm}^2 J_H^p}{\Delta^2 (2\Delta + U_{pp})^2}$$

Where $t_{pdm}$ is the π type hopping integral ($t_{pd\pi}$) and σ type hopping integral ($t_{pd\sigma}$) in SE1 and SE2, respectively. $J_H^p$ is the Hund's couplings in ligands, while $\Delta$ and $U_{pp}$ are the energy interval between involved $d$ and $p$ orbits and the onsite Coulomb interaction in $p$ orbits. Finally, for the third mechanism that happens among $t_{2g}$ and empty $e_g^*$ via a single $p$ orbital, it can be described as:[38]

$$J_{t_{2g}-e_g}^{SE} \sim -\frac{t_{pd\pi}^2 t_{pd\sigma}^2}{\Delta^2} \left( \frac{J_H^{TM}}{(2\Delta + U_{pp})^2} + \frac{J_H^{TM}}{U_{dd}^2} \right)$$

Where $J_H^{TM}$ and $U_{dd}$ are Hund's couplings and onsite Coulomb interaction in magnetic cations. Since $J_H^{TM}$ is much larger than $J_H^p$, the third ferromagnetic interaction plays the dominant role. Furthermore, such scenario can be described as $t_{2g}$-$p$-$e_g^*$, where spin-up electrons can form σ bond with $e_g^*$ and spin-down electrons will hop into $t_{2g}$ orbits via π bond simultaneously. And the whole interaction process can only be allowed for the ferromagnetic spin configuration. To conclude, for super-exchange in CrXY compounds, there are two basic hopping process denoted as $t_{2g}$-$p$ and $e_g^*$-$p$, and they are further connected by either two onsite orthogonal $p$ orbits (SE1 and SE2) or one single $p$ orbit (SE3), further forming the complete super-exchange scenario.

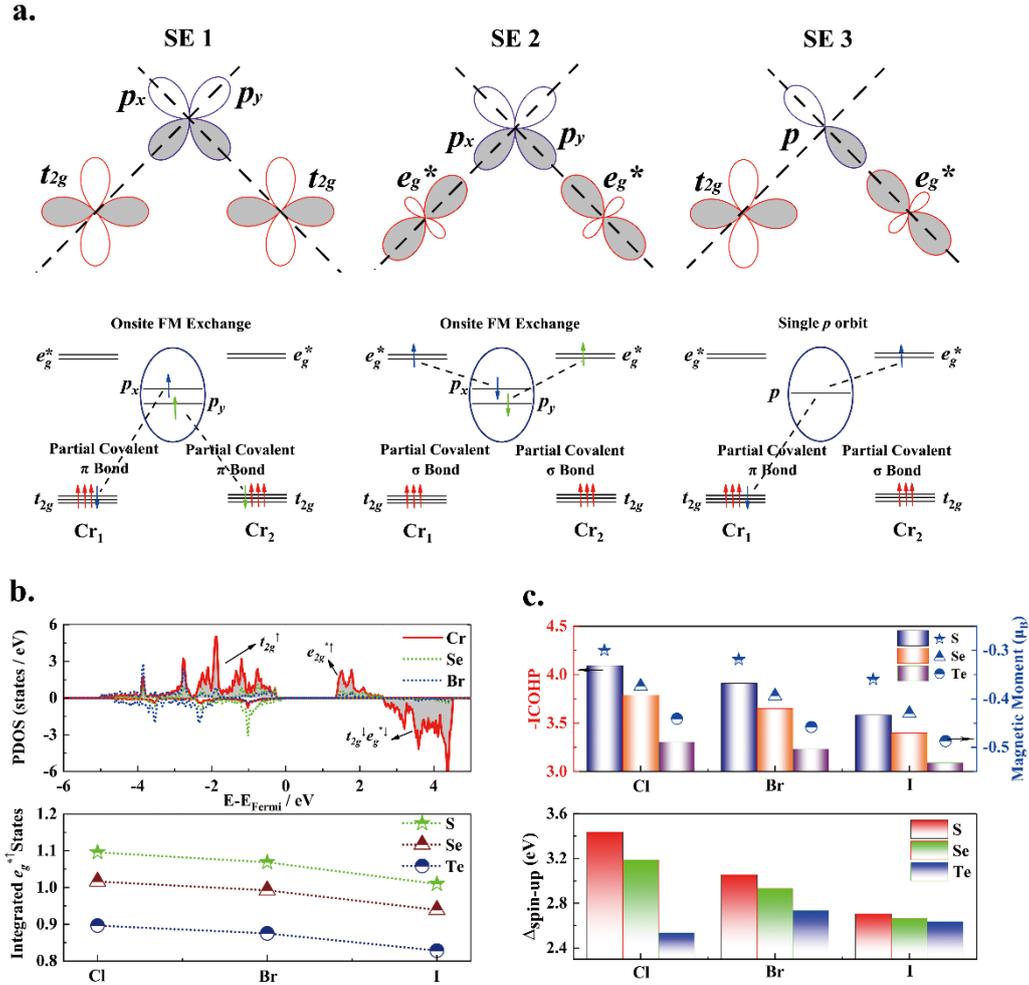

FIG. 3. a. Schematic represent of three kinds of super-exchange (SE) mechanisms as SE1, SE2 and SE3. b. Partial density of states (PDOS) for CrSeBr and the evolution of integrated $e_g^{*\uparrow}$ states. c. The evolution of -ICOHP, magnetic moment on ligands and the energy interval Δ between involved $d$ and $p$ orbits (for spin-up channel) with regard to ligands.

After identifying three basic super-exchange scenarios, the evolution of the nearest exchange integrals with regard to ligands can thus be clarified. As reflected from the above quantitative expressions, there are two major factors that determine the strength of super-exchange couplings: *d-p* hopping integral ($t_{pd\pi}$ for $t_{2g}$-*p* hopping and $t_{pd\sigma}$ for $e_g^*$-*p* hopping) and the energy interval Δ between involved *d* and *p* orbits. Firstly, for the *d-p* hopping integral, it can be reflected and described by three aspects: integrated area for unoccupied $e_g^{*\uparrow}$ and $t_{2g}^{\downarrow}e_g^{*\downarrow}$ in the partial density of states (PDOS), integrated COHP (ICOHP) for the *d-p* interaction, and the spin polarization on ligands. We firstly presented partial density of states for CrSeBr in the upper panel of Figure 3b, where occupied $t_{2g}^{\uparrow}$ and empty $e_g^{*\uparrow}$ and $t_{2g}^{\downarrow}e_g^{*\downarrow}$ are denoted as the shaded area. Varying ligands from S to Se and Te, it is found that integrated areas for unoccupied $e_g^{*\uparrow}$ and $t_{2g}^{\downarrow}e_g^{*\downarrow}$ states are decreasing (Figure 3b and Figure S7). And this trend suggests that, for heavier ligands, electrons can hop from *p* orbits into unoccupied *d* states more effectively and further lead to the partially occupied $e_g^{*\uparrow}$ and $t_{2g}^{\downarrow}e_g^{*\downarrow}$ states, thus demonstrating the benefitted hopping integral. Furthermore, the hopping process will further introduce anti-bonding characteristics which originally belong to the $e_g^{*\uparrow}$ and $t_{2g}^{\downarrow}e_g^{*\downarrow}$ states, also illustrating the observed anti-bonding peaks right below Fermi level as in the previous COHP analysis. Secondly, with the enhanced electron hopping, it can be understood that electrons can be more shared by transition metal and ligand with the higher covalency, rather than electrostatic interaction between magnetic cation and ligand anion as oxygen in conventional ionic compounds. Therefore, we quantitively evaluated the strength of covalency of *d-p* bonds by the integrated COHP (ICOHP) value. As shown in Figure 3c, bond strength gradually decreases from ionic to covalent feature, further being accompanied by the increasing $J_1$ at the same time. Thus, when evolving from S to Se and Te, the decreased electronegativity of ligand can result in such enhanced covalency. As for the third aspect, due to the more effective σ type hopping between $e_g^{*\uparrow}$ and $p^{\uparrow}$ orbits than the π type hopping between $t_{2g}^{\downarrow}$ and $p^{\downarrow}$, the spin-up electrons left on $p$ orbits would be lesser than the spin-down electrons, thus inducing spin polarization on ligands with the negative magnetic moment. As presented in the upper panel of Figure 3c, varying ligands from S to Se and Te, the magnetic moment on ligands is magnified, further reflecting the promoted electron hopping process. On the other aspect, the energy interval between involved $d$ and $p$ orbits are also gradually reduced as shown in Figure 3c (for spin up) and Figure S8 (for spin down), which originate from the weakened splitting field of ligands with lower electronegativity. Therefore, bonding and anti-bonding states as $e_g$ and $e_g^*$ are with a closer energy interval, and such reduced energy gaps are expected to benefit the super-exchange process. Huang *et al.* regulated this point from a theoretical view and further achieved the reduced gaps via doping heavy transition metals rather than controlling ligands.[39]

On the other hand, the large $J_2$ in telluride compounds can also be clarified by modeling the interacting scenario as a "cation-anion-anion-cation" system,[38] which can form right angle with each other because of the Janus monolayer geometry. There exist two kinds of interaction scenarios, which both involves the interaction between two adjacent ligands and can be denoted as super-super-exchange (SSE). As illustrated in Figure 4a and 4b, we further classify them into SSE1 and SSE2, which corresponds to $t_{2g}$-$p_1$-$p_2$-$t_{2g}$ and $e_g^*$-$p_1$-$p_2$-$e_g^*$ interaction. Similar to the former nearest exchange couplings, $t_{2g}/e_g^*$-$p$ hopping always exists, and is further connected by the formation of bonds between off-site $p$ orbits, rather than the Hund's couplings in onsite $p$ orbits. Due to the orbital symmetry matching, SSE1 and SSE2 in Figure 4a and 4b can result in the σ and π bonds among adjacent $p$ orbits respectively. Such $p$-$p$ interaction favors the opposite spin-polarization direction on ligands and further produces the AFM coupling between $t_{2g}/e_g^*$ orbits in the second nearest Cr pairs. Therefore, both mechanisms for $J_2$ all produce AFM couplings and generally exist in the Janus monolayer, and its strength is largely determined by the offsite *p-p* bonds. In our CrXY system, it is found that, varying from S to Se and Te, $J_2$ slightly decreases and reaches an anomalous large negative value in telluride system. It can be understood that, the metallic property of telluride ligand possesses the delocalized and longer interaction ability of *p* electrons, and thus the exchange among two anions can be significantly amplified. To demonstrate such mechanism, we have presented spatial charge distribution for the two-fold $e_g^*$ states at Γ point (Figure 4c), where the significant *p* orbit components can be observed due to the *d-p* hybridization. Compared to selenium, telluride compounds are with a more delocalized feature and can be more beneficial for the *p-p* interaction. Furthermore, integrated COHP for X-X and Y-Y bonds (X = Se/Te, and Y = Cl/Br) are also presented in Figure 4d. And such bonding strength among adjacent ligands can be used as a quantitative index to describe the offsite *p-p* bonds. Results show that, for ligands with tight electronic shell, the interaction among adjacent halides as Cl and Br is negligible. Therefore, Se-Se and Te-Te dominate the *p-p* couplings in SSE, and compared to selenium compounds, Te-Te possesses a larger -ICOHP value and can strengthen the anti-ferromagnetic $J_2$ more effectively.

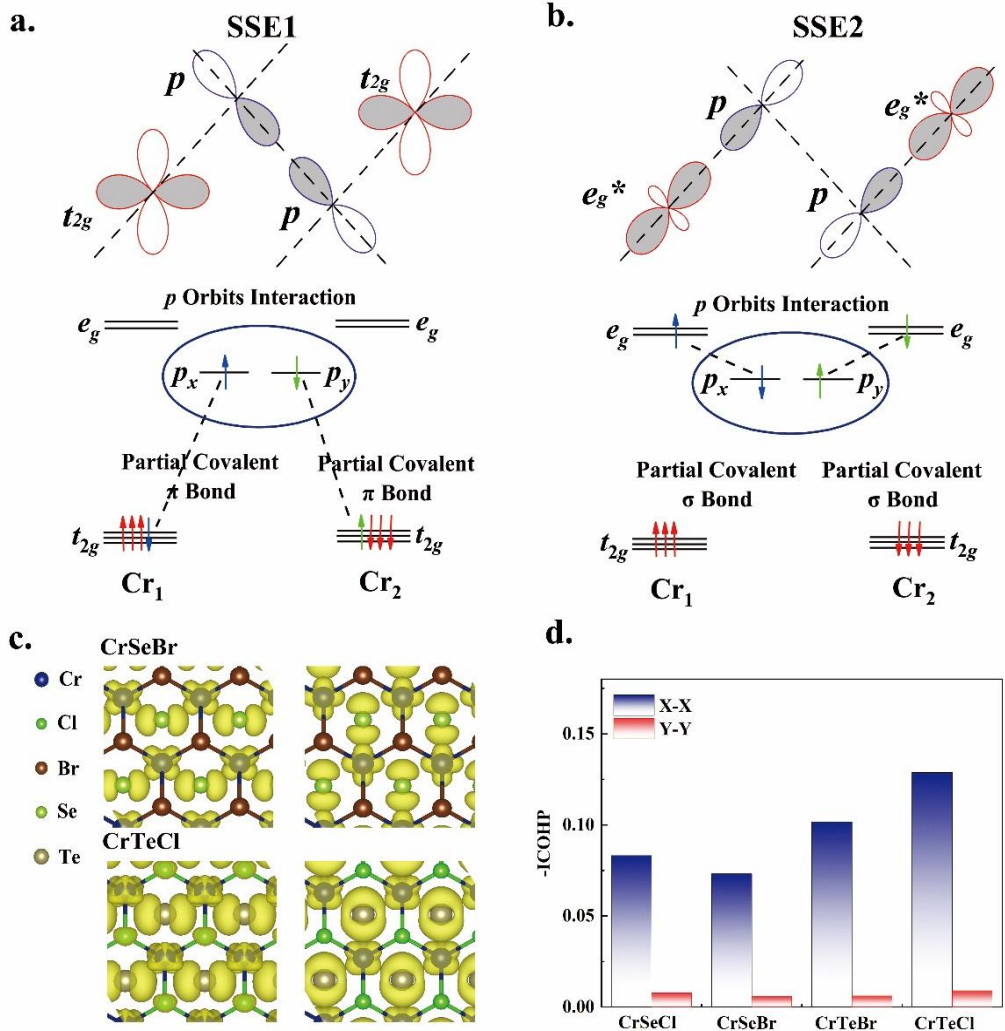

FIG. 4. a. Schematic representation of two kinds of super-super-exchange (SSE) mechanisms as a. SSE1 and b. SSE2. c. Spatial charge distribution for double degenerate $e_g^*$ orbits at $\Gamma$ point, for CrSeBr and CrTeCl respectively, where iso-surface is set to 0.01 e/Å$^3$. d. Evolution of -ICOHP for X-X and Y-Y bonds (X = Se/Te, Y = Cl/Br).

Next, we would like to give some further suggestions on achieving robust ferromagnetism in two-dimensional monolayer. Firstly, under the premise of $t_{2g}^3 e_g^{*0}$ electronic configuration, three kinds of super-exchange process will lead to ferromagnetic order, which is the general feature of this electronic structure. Furthermore, experimental achieved ferromagnetic monolayer as CrI$_3$ and CrGeTe$_3$ are both with the $t_{2g}^3 e_g^{*0}$ electronic configuration. And we can further modulate super-exchange couplings in $t_{2g}^3 e_g^{*0}$ via substituting ligands with lower electronegativity, to strengthen the *d-p* covalency and thus benefit the electron hopping process. However, heavier ligands, especially for Te, will also bring a side effect as the delocalized *p* electrons and further enhance the second-nearest interaction. To verify the credence of our theory, we examine other two widely explored semiconducting ferromagnetic systems, as $CrY_3$ (Y = Cl/Br/I) and $Cr_2Ge_2X_6$ (X = S/Se/Te), which are both with the $t_{2g}^3 e_g^{*0}$ electronic configuration.[5][6][7] The evolution of $J_1$ against ligands is plotted in Figure S9 and S10, which demonstrates that, the nearest ferromagnetic exchange integrals can both be strengthened by ligands with the lower electronegativity, being the same as the CrXY family. However, due to the different geometries, the second-nearest exchange integral will differ (ferromagnetic $J_2$ for $CrY_3$ and anti-ferromagnetic $J_2$ for $Cr_2Ge_2X_6$, being consistent with the previous literature[41][44]), but their magnitudes can both be magnified with the ligand evolution towards heavier elements (Figure S9 and S10), again verifying our theory.

IV. CONCLUSION

In conclusion, we have theoretically explored the role of ligands played in modulating super-exchange process in semiconducting ferromagnets. Based on the metallic phase of $CrX_2$ (X = S/Se/Te), a layer of original ligands is replaced by halides, which results in a series of Janus monolayer as CrSeBr, CrSeCl and CrTeBr, with robust ferromagnetic couplings. Three kinds of super-exchange paths are revealed, being ferromagnetic due to the special $t_{2g}^3 e_g^{*0}$ electronic configuration. Detailed analysis on exchange integrals shows that, the nearest ferromagnetic $J_1$ gradually increases with the ligands with the lower electronegativity. And we demonstrated that, heavier ligands can strengthen the electron hopping integral and reduce the energy interval between *d* and *p* orbits at the same time, further benefiting the super-exchange process. As for the second nearest $J_2$, its magnitude can also be enlarged due to the more delocalized *p* electrons. Therefore, we propose a fundamental understanding on the modulation of ferromagnetic couplings in low-dimensional semiconductors via ligands, serving as a theoretical guidance on the further engineering of magnetic materials.

## ACKNOWLEDGEMENTS

J. X. thanks Feng Zhi Ning and Ning Kang for generous support and helpful discussions. Q.F.Z. was supported by National Key Research and Development Program of China (No. 2017YFB0702100) and National Natural Science Foundation of China (11404017). D.L. acknowledges support by the European Regional Development Fund in the IT4Innovations national supercomputing center-Path to Exascale project, No. CZ.02.1.01/0.0/0.0/16_013/0001791 within the Operational Programme Research, Development and Education and by the Ministry of Education by Czech Science Foundation project No. 17-27790S, and grant No. 8J18DE004 of Ministry of Education, Youth, and Sport of the Czech Republic.## REFERENCE

Supporting Information

# Modulating Super-Exchange Strength to Achieve Robust Ferromagnetic Couplings in Two-Dimensional Semiconductors"

*Jiewen Xiao[1], Dominik Legut[2], Weidong Luo[3], Xiaopeng Liu[1], Ruifeng Zhang[1] and Qianfan Zhang[1*]*

1. School of Materials Science and Engineering, Beihang University, Beijing 100191, P. R. China.

2. IT4Innovations & Nanotechnology Centre, VSB-Technical University of Ostrava, 17.listopadu 2172/15, Ostrava CZ-70800, Czech Republic.

3. Institute of Natural Sciences, Shanghai Jiao Tong University, Shanghai 200240, P. R. China

*Corresponding authors: [qianfan@buaa.edu.cn](qianfan@buaa.edu.cn).*

## Section SI. Structure and Stability

Started from the 1-T $CrS_2$, $CrSe_2$ and $CrTe_2$ monolayer phase, we have replaced one layer of S/Se/Te by Cl/Br/I and further compare the energies of other common 2D structures, to select out components with the 1-T phase as the most stable structure. Here, three kinds of 2D structures are considered, which are 1-T phase, 2-H phase and the monolayer structure of the bulk phase CrOCl, as shown in Figure S1. And the most stable structure for the Cr based system with different ligand atoms is summarized in Table S1.

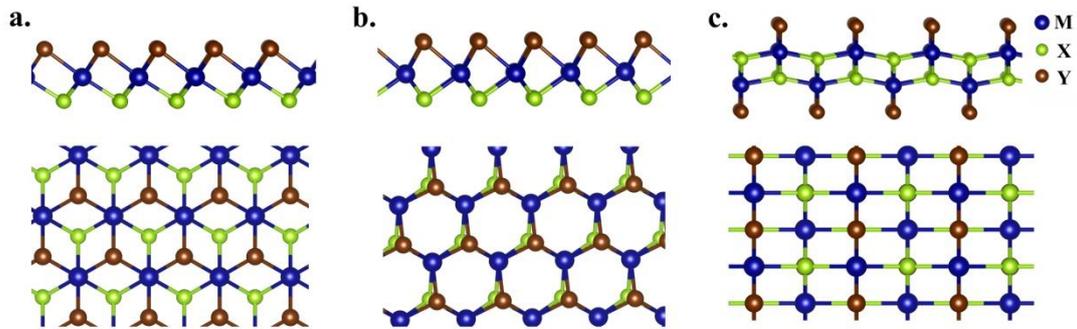

**Figure S1.** The geometric structure of a. 1-T phase, b. 2-H phase and c. CrOCl phase (the monolayer unit of bulk phase CrOCl), respectively. Here, M denotes metallic atoms while X and Y represent ligand atoms.

**Table S1.** The most stable structure for CrXY composites, where X = S/Se/Te and Y = Cl/Br/I.

| CrXY  | Phase | CrXY   | Phase | CrXY   | Phase |
|-------|-------|--------|-------|--------|-------|
| CrSCl | CrOCl | CrSeCl | 1-T   | CrTeCl | 1-T   |
| CrSBr | CrOCl | CrSeBr | 1-T   | CrTeBr | 1-T   |
| CrSI  | CrOCl | CrSeI  | CrOCl | CrTeI  | CrOCl |

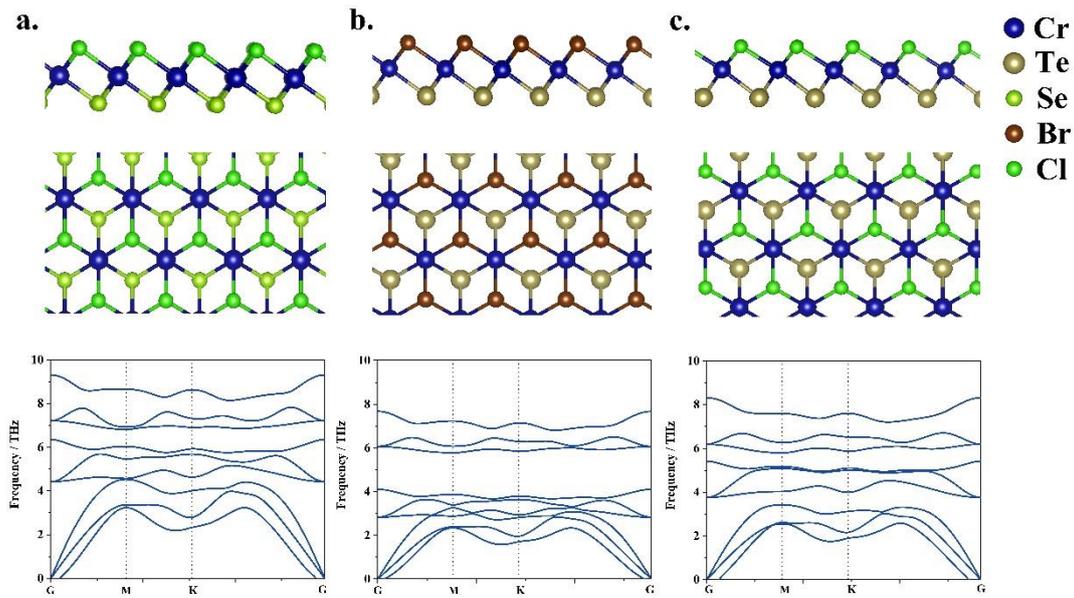

**Figure S2.** The geometric structure and the corresponding phonon spectra of 1-T a. CrSeCl, b. CrTeBr, c. CrTeCl respectively, where side view, top view and phonon spectra are presented from top to bottom.

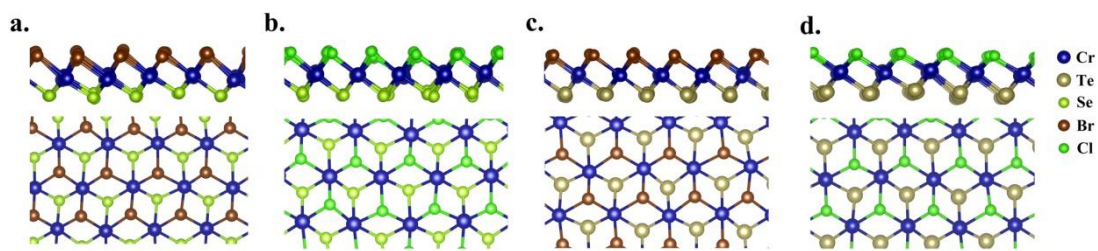

**Figure S3.** Atomic configurations for a. CrSeCl, b. CrSeBr, c. CrTeBr and d. CrTeCl after 20ps AIMD simulation at 300K.

# Section SII. Magnetic Properties

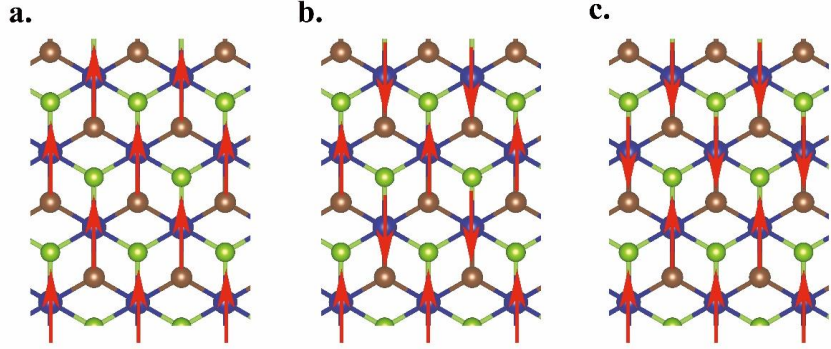

**Figure S4.** Three kinds of magnetic configurations for 1-T phase CrXY (where X and Y refer to different ligands): a. ferromagnetic order; b. c. anti-ferromagnetic orders.

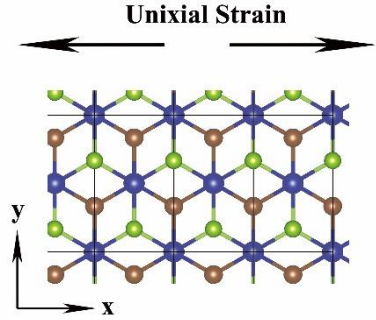

**Figure S5.** Schematic representation of the applied uniaxial strain.

**Table S2.** Magnetic anisotropic energy (MAE) for CrSeCl, CrSeBr and CrTeBr, when ± 1.5% uniaxial strain is applied in the $x$ direction (shown in Figure S5), and MAE is defined as $E_x - E_y$.

|  | $MAE_{CrSeCl}/\mu eV$ | $MAE_{CrSeBr}/\mu eV$ | $MAE_{CrTeBr}/\mu eV$ |
|---|---|---|---|
| **1.5%** | 24.04 | 19.84 | 156.29 |
| **-1.5%** | -13.58 | -25.43 | -206.96 |

**Table S3.** Exchange integrals for CrXCl ( X = S/Se/Te)

|  | CrSCl | CrSeCl | CrTeCl |
|---|---|---|---|
| $J_1$ / meV | 25.14 | 28.23 | 33.49 |
| $J_2$ / meV | -1.88 | -5.51 | -17.83 |

**Table S4.** Exchange integrals for CrXBr ( X = S/Se/Te)

|  | **CrSBr** | **CrSeBr** | **CrTeBr** |
|---|---|---|---|
| $J_1$ / meV | 25.61 | 29.48 | 34.93 |
| $J_2$ / meV | -1.54 | -4.53 | -14.62 |

**Table S5.** Exchange integrals for CrXI ( X = S/Se/Te)

|  | **CrSI** | **CrSeI** | **CrTeI** |
|---|---|---|---|
| $J_1$ / meV | 23.42 | 29.41 | 36.27 |
| $J_2$ / meV | -1.32 | -3.40 | -10.40 |

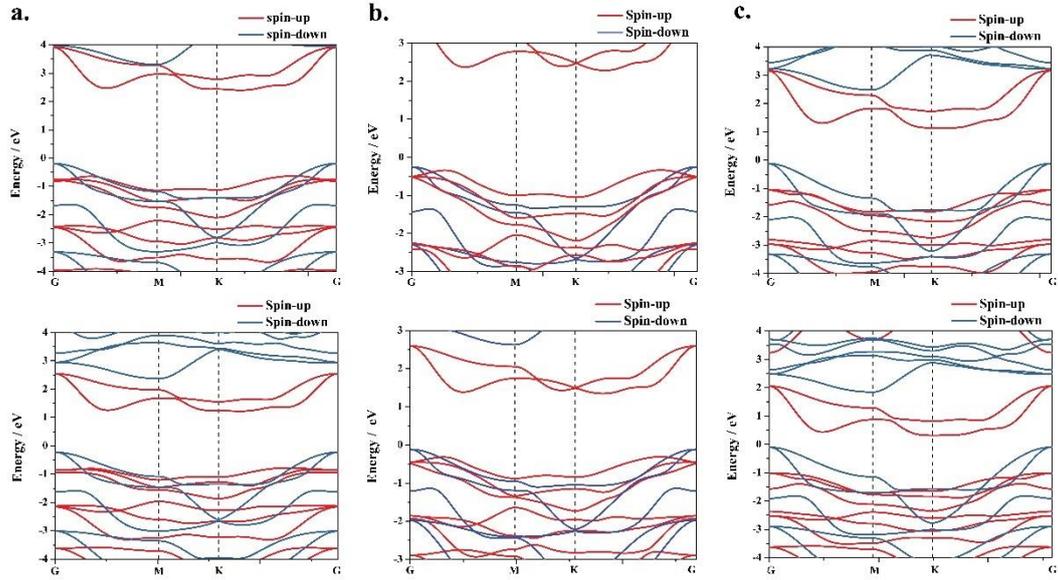

**Figure S6.** Calculated band structure based on HSE06 (upper panel) and PBE+U (lower panel) exchange-correlation scheme for a. CrSeCl, b. CrSeBr and c. CrTeBr, respectively

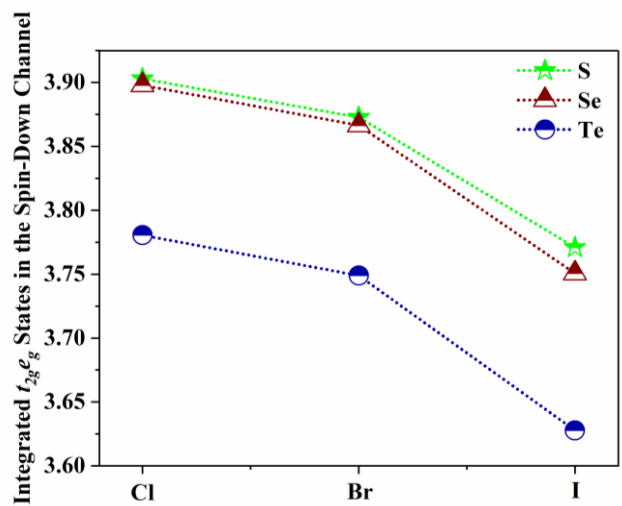

**Figure S7.** The evolution of integrated $t_{2g}^\downarrow e_g^{*\downarrow}$ states with regard to ligands.

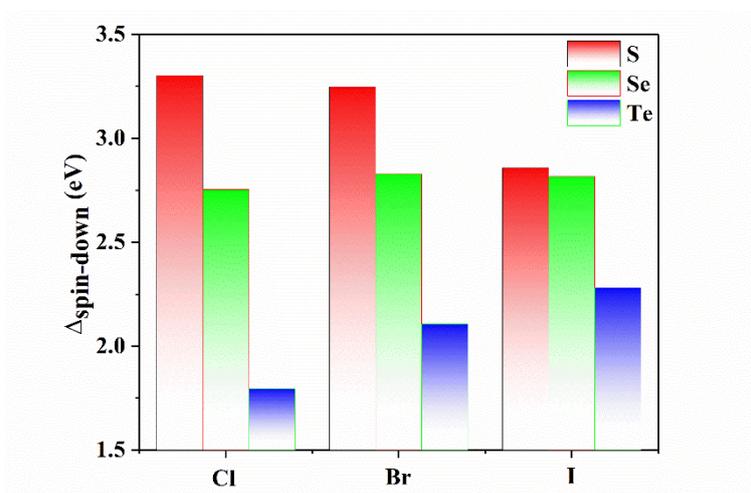

**Figure S8.** The evolution of the energy gap $\Delta$ between involved $d$ and $p$ orbits.

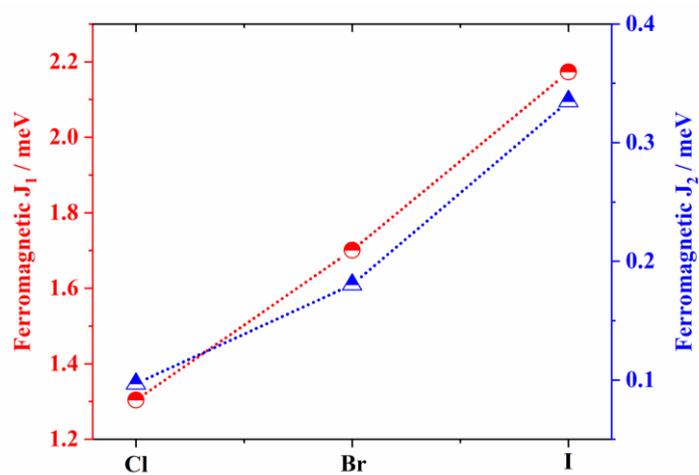

**Figure S9.** The evolution of the nearest and the second nearest exchange integral for $CrY_3$ (Y = Cl/Br/I).

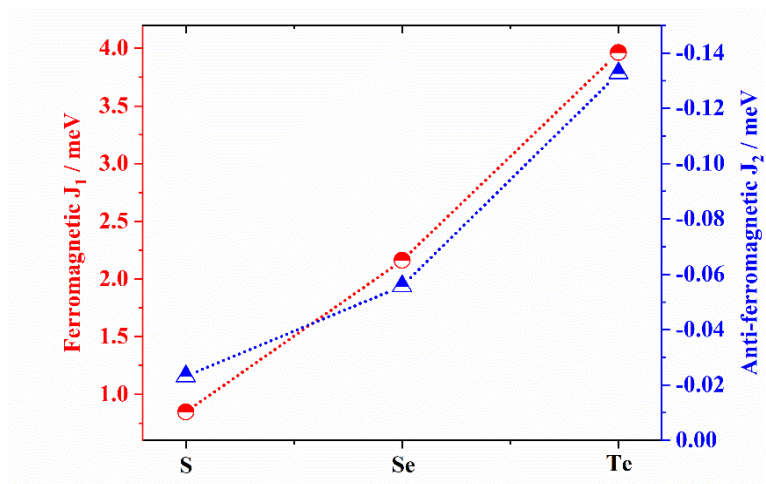

**Figure S9.** The evolution of the nearest and the second nearest exchange integral for $Cr_2Ge_2X_6$ (X = S/Se/Te).